\documentstyle[12pt]{article}
\def\dspace{\baselineskip=0.3 in}
\begin{document}
\dspace

{\large

\centerline{\bf Future universe with ${\rm w} < - 1$ Without Big Smash} }

\vspace{0.5cm}

\centerline{\bf S.K.Srivastava}

\centerline{ Department of Mathematics,}

\centerline{ North Eastern Hill University,}

\centerline{ Shillong - 793022, India } 

\centerline{\bf  e - mail :srivastava@nehu.ac.in}

\vspace{2cm}
\centerline{\bf Abstract}

It is demonstrated that if cosmic $dark$ $energy$ behaves like a fluid
with equation of state $p = {\rm w} \rho$ ($p$ and $\rho$ being pressure and
energy density respectively) as well as generalized Chaplygin gas
simultaneously,  Big Rip or Big Smash problem does not arise even for equation
of state parameter ${\rm w}< -1$. Unlike other phantom models, here, the scale
factor for the future universe is found regular for all time. 

\vspace{2cm}

Experimental probes, during last few years suggest that the present universe is spatially flat as well as it is dominated by yet unknown form of $dark$ $energy$ [1,2]. Moreover, studies of Ia Supernova [3,4] and WMAP [5,6] show accelerated expansion of the present universe such that ${\ddot a} > 0$ with $a(t)$ being the scale factor of the Friedmann Robertson Walker line-element

$$ dS^2 = dt^2 - a^2(t)[dx^2 + dy^2 + dz^2]. \eqno(1)$$

Theoretically accelerated expansion of the universe is obtained when the cosmological model is supposed to be dominated by a fluid obeying the equation of state(EOS) $ p ={\rm w} \rho$ with $p$ as isotropic pressure, $\rho$ as energy density and $ - 1 \le {\rm w} < - 1/3$.

In the recent past, it was pointed out that the current data also allowed
${\rm w}< -1$ [7]. Rather, in refs.[8,9,10], it is discussed that these data
favor ${\rm w}< -1$ being EOS parameter for $phantom$ $dark$ $energy$. Analysis of recent Ia Supernova data also support ${\rm w}< -1$ strongly [11,12,13].

Soon after, Caldwell [8] proposed the $phantom$ $dark$ $energy$ model
exhibiting cosmic doomsday of the future universe, cosmologists started making
efforts to avoid this problem using  ${\rm w} <
-1$ [14,15]. In the braneworld scenario, Sahni and Shtanov has obtained
well-behaved expansion of the future universe without Big Rip problem with
${\rm w} < -1$. They have shown that acceleration is a transient phenamenon in
the current universe and the future universe will re-enter matter-dominated
decelerated phase [16]. 

It is found that GR(general relativity)-based phantom model encounters
``sudden future singularity'' leading to divergent scale factor $a(t)$, energy
density and pressure at finite time $t = t_s$. Thus the classical approach to
phantom model yields big-smash problem. For models with ``sudden future
singularity'' Elizalde, Nojiri and Odintsov [17] argued that, near $t = t_s$,
curvature invariants become very strong and energy density is very high. So,
quantum effects should be dominant for $|t_s - t| <$ one unit of time, like
early universe. This idea is pursued in refs.[18, 19,20] and it is shown that
an escape from the big-smash is possible on making quantum corrections to
energy density $\rho$ and pressure $p$ in Friedmann equations. 

In the framework of Robertson-Walker cosmology, Chaplygin gas (CG) is also
considered as a good source of dark energy for having negative pressure, given
as 

$$p = - \frac{A}{\rho}  \eqno(2)$$
with $A > 0.$ Moreover, it is the only gas having supersymmetry generalization
[21,22]. Bertolami $et$ $al$ [12] have found that generalized Chaplygin gas (GCG) is better fit for latest
Supernova data. In the case of GCG, eq.(2) looks like   

$$p = - \frac{A}{\rho^{1/\alpha}},  \eqno(3)$$
where $1 \le \alpha < \infty. \alpha = 1$ corresponds to eq.(2).

In this letter,  a different prescription for GR-based  future universe, dominated
by the $dark$ $energy$ with ${\rm w} < -1$, is proposed which is not leading
to the catastrophic situations mentioned above. The scale factor, obtained
here,does not possess future singularity. In the present model, it is assumed
that the $dark$ $energy$ behaves like GCG, obeying eq.(3) as well as fluid with equation of state 
$$ p = {\rm w} \rho \quad {\rm with} \quad {\rm w} < -1 \eqno(4)$$
 simultaneously.

Connecting eq.(3) with the hydrodynamic equation

$$ {\dot \rho} = - 3 \frac{\dot a}{a} (\rho + p) \eqno(5)$$
and integrating, it is obtained that

$$\rho^{(1 + \alpha)/\alpha}(t) = A + (\rho^{(1 + \alpha)/\alpha}_0 - A) (a_0/a(t))^{3(1 + \alpha)/\alpha} \eqno(6)$$
with $\rho_0 = \rho (t_0)$ and $ a_0 = a(t_0)$, where $t_0$ is the present time. 

Eqs.(3) and (4) yield  $\rm w$ as

$$ {\rm w}(t) = - \frac{A}{\rho^{(1 + \alpha)/\alpha}(t)} \eqno(7a)$$
So, evaluation of eq.(7a) at $t = t_0$ leads to
$$ A = - {\rm w}_0 \rho^{(1 + \alpha)/\alpha}. \eqno(7b)$$
with ${\rm w}_0 = {\rm w}(t_0).$ From eqs.(6) and (7), it is obtained that

$$\rho = \rho_0 \Big[ - {\rm w}_0 + (1 + {\rm w}_0)(a_0/a(t))^{3(1 + \alpha)/\alpha} \Big]^{\alpha/(1 + \alpha)} \eqno(8)$$
 with ${\rm w}_0 < - 1$.

In the homogeneous model of the universe, a scalar field $\phi(t)$ with
potential $V(\phi)$ has energy density
$$\rho_{\phi} = \frac{1}{2} {\dot \phi}^2 + V(\phi) \eqno(9a)$$
and pressure
$$ p_{\phi} = \frac{1}{2} {\dot \phi}^2 - V(\phi). \eqno(9b)$$

Using eqs.(3), (4), (7) and (8), it is obtained that

$${\dot \phi}^2 = \frac{\rho^{(1 + \alpha)/\alpha} + \rho^{(1 + \alpha)/\alpha}_0 {\rm w}_0}{\rho}. \eqno(10)$$

Connecting eqs.(8) and (10), it is obtained that

$${\dot \phi}^2 = \frac{( 1 + {\rm w}_0)\rho^{(1 + \alpha)/\alpha}_0 (a_0/a)^{3(1 + \alpha)/\alpha}}{[- {\rm w}_0 + ( 1 + {\rm w}_0)(a_0/a)^{3(1 + \alpha)/\alpha}]^{\alpha/(1 + \alpha)}}  . \eqno(11)$$
This equation shows that ${\dot \phi}^2 > 0$ (giving positive kinetic energy)for ${\rm w}_0 > -1$, which is
the case of quintessence  and ${\dot \phi}^2 < 0$ (giving negative kinetic energy)for ${\rm w}_0 < -1$, being
the case of super-quintessence (phantom).  As a reference, it is relevant to
mention that, long back, Hoyle and Narlikar used C - field (a scalar called creation field) with
negative kinetic energy for steady-state theory of the universe [23].

Thus, it is shown that dual bahaviour of dark energy fluid, obeying eqs.(3)
and (4) is possible for scalars, frequently used for cosmological dyanamics. So, this assumption is not unrealistic.

Now the Friedmann equation, with dominance of $dark$ $energy$ having double
fluid behaviour , is

$$ \Big(\frac{\dot a}{a} \Big)^2 = H^2_0 \Omega_0 \Big[ |{\rm w}_0| + (1 - |{\rm w}_0|)(a_0/a(t))^{3(1 + \alpha)/\alpha} \Big]^{\alpha/(1 + \alpha)}, \eqno(12a)$$
where $|{\rm w}_0| > 1.$  $H_0$ is the present value of Hubble's constant and
$\Omega_0 = \rho_0 / \rho_{{\rm cr},0}$ with $\rho_{{\rm cr},0} = 3 H_0^2/ 8
\pi G$ ($G$ being the Newtonian gravitational constant).

Neglecting higher powers of $\frac{1 - |{\rm w}_0|}{|{\rm
    w}_0|}(a_0/a(t))^{3(1 + \alpha)/\alpha}$ , eq.(12a) is written as

$$\frac{\dot a}{a} \simeq H_0 \sqrt{\Omega_0} |{\rm w}_0|^{\alpha/2(1 +
  \alpha)}\Big[1 + \frac{\alpha (1 - |{\rm w}_0|)}{2(1 + \alpha)|{\rm
    w}_0|}(a_0/a(t))^{3(1 + \alpha)/\alpha} \Big] \eqno(12b)$$

Eq.(12b) is integrated to
 
$$ a(t) = \frac{a_0 }{[2(1 + \alpha)|{\rm w}_0|]^{\alpha/3(1 + \alpha)}} \Big[ (\alpha
+ 2(1 + \alpha) |{\rm w}_0|) e^{ 6H_0 |{\rm w}_0|^{\alpha/2(1 + \alpha)}
  \sqrt{\Omega_0} (t - t_0)} - \alpha (1 - |{\rm w}_0|) \Big]^{\alpha/3(1 + \alpha)}. \eqno(13)$$
yielding accelerated expansion of the universe with $a(t) \to \infty$ as $ t
\to \infty$ , supporting observational evidences of Ia Supernova [3,4] and
WMAP [5,6]. It is interesting to see that expansion ,obtained here, is free
from ``finite time future singularity'' unlike other GR-based phantom
models. It is due to GCG behaviour of phantom dark energy.

Moreover, eq.(8) and (13) that energy density grows with time for ${\rm w}_0 <
-1$ and decreases for ${\rm w}_0 > -1.$ Also $\rho \to \rho_0 |{\rm
  w}_0|^{\alpha/3(1 + \alpha)} $ (finite) and $p \to - p_0/ |{\rm
  w}_0|^{\alpha/3(1 + \alpha)} $ as $t \to \infty$. Eqs.(7) and (8) imply
time-dependence of EOS parameter 

$$ {\rm   w} = - |{\rm   w}_0|[ |{\rm   w}_0| - (|{\rm   w}_0| -
1)(a_0/a(t))^{3(1 + \alpha)/\alpha}]^{-1} \eqno(14)$$
with $a(t)$, given by eq.(13). This equation shows that ${\rm   w} \to - 1$ asymptotically.

The horizon distance for this case ($a(t)$ given by eq.(16)) is obtained as

$${ d_H(t)} \simeq \frac{3(1 + \alpha)a(t)}{\alpha a_0} \Big[\frac{2(1 +
  \alpha)|{\rm   w}_0|}{\alpha + (\alpha + 2)|{\rm   w}_0|} \Big]^{\alpha/3(1 + \alpha)}exp[6H_0 |{\rm w}_0|^{\alpha/2(1 + \alpha)}
  \sqrt{\Omega_0} (\alpha t/3(1 + \alpha))] \eqno(15a)$$
showing that 

$${ d_H(t)} > a(t) . \eqno(15b)$$
So, horizon grows more rapidly than the scale factor impling colder and darker universe. It is like flat or open universe without dominance of dark energy.

In this case, Hubble's distance is

$${ H^{-1}} = \frac{3(1 + \alpha))}{\alpha H_0 \sqrt{\Omega_0} |{\rm
  w}_0|^{ \alpha/2(1 + \alpha)}} \Big\{1 - \frac{\alpha (1 - |{\rm
  w}_0)|}{\alpha +   (\alpha + 2)|{\rm   w}_0|}exp[-H_0 |{\rm w}_0|^{\alpha/2(1 +  \alpha)} \sqrt{\Omega_0}(t - t_0)] \Big\} \eqno(16)$$
showing  its growth with time such that ${ H^{-1}} \to \frac{3(1 + \alpha))}{\alpha H_0 \sqrt{\Omega_0}} |{\rm
  w}_0|^{- \alpha/2(1 + \alpha)} \ne 0$ as $t \to \infty$. Here, $H^{-1}_{\infty}$
  is found large and finite. It
  means that, in the present case,  galaxies will not disappear when $t \to
  \infty$. It is unlike  $phantom$ models with future singularity expanding as
  $|t - t_s|^n$ for $n < 0$ , where
  galaxies are expected to vanish near future singularity time $t_s$ [8] as
  $H^{-1} \to 0$ for $t \to t_s$. In Barrow's model [24]
 
$$H^{-1} = \frac{B + C t^q + D (t_s -t)^n}{q C t^{q-1} - D n (t_s -
  t)^{n-1}},\eqno(17)$$
where $ B, C, D$ are positive constants and $ q >0.$ Eq.(17) shows that, for
  $n< 1, H^{-1} \to 0$ as $t \to t_s$ and at $t = t_s, H^{-1}$ is finite for $n >1$ .In the model, taken by Nojiri and Odintsov [18]

$$H^{-1} = [ {\tilde H(t)} + A^{\prime} |t_s - t|^n]^{-1}, \eqno(18)$$
where ${\tilde H(t)}$ is a regular function of $t$ and $ A^{\prime} > 0$. This
equation shows that, for $n < 0, H^{-1} \to 0$ as $t \to t_s$ and it is finite at
$t = t_s$ for $n >0$ .

Thus, it is found that if phantom fluid behaves like GCG and fluid with $p =
{\rm w} \rho$, it is possible to get accelerated growth of  scale factor
of the future universe for time $t_0 < t < \infty$ with no future singularity
contrary to other phantom models. Here also, it is obtained that energy
density and pressure increase with time , asymptotically approaching finite
values $\rho_0 |{\rm w}_0|^{\alpha/3(1 + \alpha)} > \rho_0$ and $ - p_0/|{\rm
  w}_0|^{1/3(1 + \alpha)} > - p_0$ respectively. It is unlike GR-based models, driven by EOS $p =
{\rm w} \rho$, with ${\rm w} < -1$ having future singularity at $t = t_s$,
where $\rho$ and $p$ are divergent [8,14] or $\rho$ is finite and $p$ is
divergent [18,24]. Based on Ia Supernova data, Singh $et$ $al$ [13] have
estimated ${\rm w}_0$ for models in the range $ -2.4 < {\rm w}_0 < - 1.74$
upto $95 \%$ confidence level. Taking this estimate as an example with $\alpha
=3$, $\rho_{\infty} = \rho (t \to \infty)$ is found in the range $ 1.15 \rho_0
<\rho_{\infty} < 1.24 \rho_0$. This does not yield much increase in $\rho$
as $t \to \infty$. But if this model is realistic and future experiments
support large $|{\rm w}_0|$, $\rho_{\infty}$ will be very high. In both cases,
small or large values of $|{\rm w}_0|$, increase in $\rho$ indicates creation
of phantom dark energy in future. It may be due to decay of some other
components of energy in universe, which is not dominating, for example cold
dark matter.

It is interesting to see that big-smash problem does not arise in the present
model. In refs.[17,18,19,20], for models with future singularity, escape from
cosmic doomsday is demonstrated using quantum corrections in field equations
near $t=t_s$. Here, using classical approach, a model for phantom
cosmology, with accelerated expansion, is explored which is free from catastrophic situations. This model is
derived from Friedmann equations using the effective role of GCG behaviour in
a natural way.

\centerline{\bf References}

\bigskip

\noindent [1] A.D. Miller $et$ $al$ , Astrophys. J. Lett. {\bf 524} (1999) L1; P. de Bernadis $et$ $al$ , Nature (London){\bf 400} (2000) 955; A.E. Lange $et$ $al$ , Phys. Rev.D{\bf 63} (2001) 042001; A. Melchiorri $et$ $al$ , Astrophys. J. Lett. {\bf 536} (2000) L63.

\bigskip

\noindent [2] S. Hanay $et$ $al$ , Astrophys. J. Lett. {\bf 545} (2000) L5.

\bigskip

\noindent [3] S. Perlmutter $et$ $al$ , Astrophys. J. {\bf 517} (1999) 565.

\bigskip

\noindent [4] A.G. Riess $et$ $al$ , Astron. J. {\bf 116} (1998)1009.

\bigskip

\noindent [5] D.N. Spergel $et$ $al$, astro-ph/0302209.

\bigskip

\noindent [6] L. Page $et$ $al$, astro-ph/0302220.

\bigskip

\noindent [7] V.Faraoni, Phys. Rev.D{\bf 68} (2003) 063508; R.A. Daly $et$ $al$, astro-ph/0203113; R.A. Daly and E.J. Guerra, Astron. J. {\bf 124} (2002) 1831;  R.A. Daly, astro-ph/0212107; S. Hannestad and E. Mortsell, Phys. Rev.D{\bf 66} (2002) 063508; A. Melchiorri $et$ $al$, Phys. Rev.D{\bf 68} (2003) 043509; P. Schuecker $et$ $al$, astro-ph/0211480.

\bigskip
\noindent [8]  R.R. Caldwell, Phys. Lett. B {\bf 545} (2002) 23; R.R. Caldwell, M. Kamionkowski and N.N. Weinberg, Phys. Rev. Lett. {\bf 91} (2003) 071301.

\bigskip

\noindent [9]  H. Ziaeepour, astro-ph/0002400 ; astro-ph/0301640;
P.H. Frampton and T. Takahashi,  Phys. Lett. B {\bf 557} (2003) 135;
P.H. Frampton, hep-th/0302007;  S.M. Carroll $et$ $al$, Phys. Rev.D{\bf 68}
(2003) 023509 ; P.Singh, gr-qc/0502086 .

\bigskip

\noindent [10] J.M.Cline $et$ $al$, hep-ph/0311312.

\bigskip

\noindent [11] U. Alam , $et$ $al$, astro-ph/0311364 ; astro-ph/0403687.

\bigskip

\noindent [12] O. Bertolami, $et$ $al$, MNRAS, {\bf 353}, (2004) 329 [astro-ph/0402387].

\bigskip

\noindent [13] P. Singh, M. Sami $\&$ N. Dadhich, Phys. Rev.D{\bf 68} (2003) 023522; M. Sami $\&$ A. Toporesky, gr-qc/0312009. .
\bigskip

\noindent [14] B. McInnes, JHEP, {\bf 08} (2002) 029; hep-th/01120066.

\bigskip

\noindent [15] Pedro F. Gonz$\acute a$lez-D$\acute i$az, Phys. Rev.D{\bf 68}
(2003) 021303(R); V.K.Onemli $et$ $al$, Class. Quan. Grav. {\bf 19} (2002)
       4607 (gr-qc/0204065); Phys.Rev. {\bf D 70} (2004) 107301 (gr-qc/0406098);Class. Quan. Grav. {\bf 22} (2005) 59 (gr-qc/0408080) .

\bigskip

\noindent [16] V. Sahni $\&$ Yu.V.Shtanov, JCAP {\bf 0311} (2003) 014;
astro-ph/0202346 ; G.Calcagni, Phys. Rev.D{\bf 69} (2004) 103508 ; V. Sahni, astro-ph/0502032.

\bigskip

\noindent [17] E. Elizalde, S. Noriji and S. D. Odintsov ,Phys. Rev.D{\bf 70}
(2004) 043539 [hep-th/0405034].

\bigskip

\noindent [18] S. Nojiri and S. D. Odintsov , Phys. Lett. {\bf B 595} (2004) 1
[hep-th/0405078] ; Phys. Rev.D{\bf 70} (2004) 103522 [hep-th/0408170].

\bigskip

\noindent [19] S.K.Srivastava, hep-th/0411221.

\bigskip

\noindent [20] S. Nojiri, S. D. Odintsov and S. Tsujikawa , hep-th/0501025.

\bigskip

\noindent [21]  R. Jackiw, `` (A particle field theorist's) Lecture on Supersymmetric Non-Abelian Fluid Mechanics and d-branes'', physics/0010042.

\bigskip

\noindent [22] M. C. Bento, O. Bertolami, A.A.Sen, Phys. Rev. {\bf D 66}
(2002)043507 [gr-qc/0202064]; N. Bilic, G.B.Tupper and R. Viollier,
      Phys. Lett. B {\bf 535} (2002) 17; J.S. Fabris, S.V.Goncalves and
      P.E. de Soyza, astro-ph/0207430; V. Gorini, A. Kamenshchik and
      U. Moschella, Phys. Rev. {\bf D 67} (2003) 063509 [astr0-ph/0210476];
      C. Avelino, L.M.G. Beca, J.P.M. de Carvalho, C.J.A.P. Martins and P. Pinto Phys. Rev. {\bf D 67} (2003) 023511 [astr0-ph/0208528].

\bigskip

\noindent [23] F. Hoyle and J.V.Narlikar, MNRAS {\bf 108} (1948) 372; {\bf
  109} (1949) 365; Proc. Roy. Soc. {\bf A 282} (1964) 191; MNRAS {\bf 155}
  (1972) 305; J.V. Narlikar and T.Padmanabhan  Phys. Rev. {\bf D 32}
(1985) 1928;  F. Hoyle , G. Burbidge and  J.V.Narlikar, $A$ $Different$ $Approach$ $to$ $Cosmology$ (Cambridge University Press, Cambridge, England, 2000).

\bigskip

\noindent [24]J.Barrow, Class. Quan. Grav. {\bf 21} (2004) L79 - L82 {gr-qc/0403084].

\end{document}